\documentclass[prl, aps, twocolumn, groupedaddress, superscriptaddress]{revtex4}
\usepackage{amsmath}
\usepackage{amsfonts}
\usepackage{amssymb}
\usepackage{amsmath}
\usepackage{amssymb}
\usepackage{epsfig}
\usepackage{graphicx}
\usepackage{hyperref}
\usepackage{subfigure}

\begin{document}
\preprint{PRL}
\title{Plasmon wakefields and dispersive properties of metallic nanostructures}
\author{S. Ali}
\affiliation{National Centre for Physics, Shahdra Valley Road, Quaid-i-Azam University Campus, Islamabad 44000, Pakistan.}
\email{shahid\_gc@yahoo.com}
\author{H. Ter\c cas}
\affiliation{CFIF, Instituto Superior T\'{e}cnico, Av. Rovisco Pais 1, 1049-001Lisboa, Portugal.}
\author{J. T. Mendon\c ca}
\affiliation{CFIF, Instituto Superior T\'{e}cnico, Av. Rovisco Pais 1, 1049-001Lisboa, Portugal.}
\affiliation{IPFN, Instituto Superior T\'{e}cnico, Av. Rovisco Pais 1, 1049-001Lisboa, Portugal.}

\begin{abstract}

We investigate the excitation of electrostatic wakefields in metallic nanostructures (nanowires) due to the propagation of a short electron pulse. For that purpose, a dispersive (nonlocal) dielectric response of the system is considered, accounting for both the finiteness of the system and the quantum (Bohm) difraction of the conduction electron band, generalizing the results obtained previously in the literature [Phys. Rev. Lett. \textbf{103}, 097403 (2009)]. We discuss on the stability conditions of wakefields and show that the underling mechanism can be useful to investigate new sources of radiation in the extreme-ultra-violet (XUV) range. 
\end{abstract}

\maketitle

Experimental techniques \cite{Willets} have been developed for the fabrication of metallic nanostructures (nanowires) of the order of $10$ \textrm{nm} or
less, recently receiving a considerable attention of the scientific community. Nanowires, compared to other low-dimensional systems, have two conned directions, still
leaving one unconned direction for electrical conduction. Due to their unique density of states, such systems are expected to exhibit significantly different optical, electrical, and magnetic properties from their bulk 3D crystalline counterparts, specially in the limit of small diameters. Even if the number of electrons involved in the relevant features is high, and therefore a continuum description is expected to be adequate, the current models still lack of completeness. One of the most striking cases is related with the dielectric response. It was experimentally shown that anomalous absorption can occur in thin metal films \cite{Anderegg} due to the excitation of plasmons \cite{Jones}. Liu et al. \cite{Liu} have recently approached this problem for arbitrarily shaped nanostructures, though neglecting the wavenumber dependence in their model. Recently, McMahon et al \cite{McMahon} have introduced the effects of the nonlocal response by adding dispersion terms (proportional to the wavevector $\mathbf{k}$) in the Drude dielectric function of the bulk (conduction) electrons, as described by

\begin{equation}
\epsilon({\mathbf{k},\omega})=\epsilon_{\infty}-\frac{\omega_{p}^2}{\omega(\omega+i\gamma)-v_{F}^2k^2},
\label{drude}
\end{equation}  
where $\epsilon_{\infty}(\approx 1)$ is the value for $\omega\rightarrow\infty$, $\omega_{p}$ and $v_{F}$ stand for the plasma frequency and the electron Fermi velocity, respectively, and $\gamma$ represents the electron collision frequency. However, it is well known that further quantum mechanical effects \cite {Manfredi} are associated with electrons at nanoscales, playing a significant role in the dispersion of collective modes and instabilities, and therefore are expected to play an important role in nanowires, as well. In particular, such quantum effects become important when the thermal wavelength is comparable to the typical dimensions of the system.\par
In this Letter, we extend the dispersive Drude model in order to cast these quantum corrections and derive a nonlocal dielectric constant for quasi one-dimensional nanostructures (nanowires). More specifically,  we take into account the effects of i) the finiteness of the system along the transverse direction and ii) the quantum diffraction (Bohm potential) due to a gradient of the electronic density. We then apply our result to investigate the excitation of wakefields due to the propagation of a finite electron pulse. It is shown that the features of the wakefields are intrinsically connected with the dispersive response of the system. We suggest that, due to the competition between two spatial scales (say classical and quantum), the excitation of wakefields in such materials can also lead to the emission of radiation in the XUV range of frequency.\par

\begin{figure}
\includegraphics[scale=0.75]{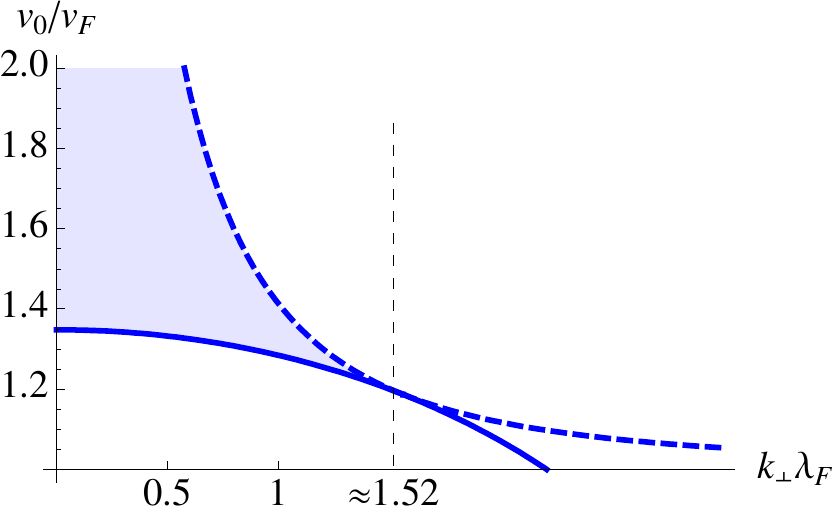}
\caption{Wakefield stability diagram for Au nanowires ($H\approx 0.817$). The two curves coincide at $k_{\perp}^{cr}\approx 1.52/\lambda_{F}$ (vertical line). The shadowed area is stable.}
\label{fig:stability}
\end{figure}

We start from a set of quantum hydrodynamical (QHD) equations,  

\begin{equation}
\frac{\partial n}{\partial t}+\boldsymbol{\nabla }\cdot \left( n\mathbf{u}\right) =0,  \label{cont1}
\end{equation}

\begin{equation}
\left(\frac{\partial }{\partial t}+i\gamma \right)\mathbf{u}+\mathbf{u}\cdot \boldsymbol{\nabla }\mathbf{u}=\frac{e}{m_{e}}\boldsymbol{\nabla }\phi -\frac{\boldsymbol{\nabla }P_{F}}{m_{e}n}+\frac{\mathbf{F}_Q}{m_{e}},  \label{momentum1}
\end{equation}

\begin{equation}
\nabla ^{2}\phi =\frac{e}{\epsilon _{0}}\left(n-n_{0}\right) ,  \label{poisson}
\end{equation}
where $n,$ $\mathbf{u},$ and $\phi $ are the electron mean density, the electron velocity, and the electrostatic potential, respectively. The closure of the system is established via an equation of state for the electrons of the conduction band

\begin{equation}
P_{F}=\frac{m_{e}v_{F}^{2}}{3n_{0}^{2}}n^{3}. \label{state}
\end{equation}
The last term in Eq. (\ref{momentum1}) corresponds to the so-called quantum (or Bohm) force and casts the effects of the quantum diffraction

\begin{equation}
\mathbf{F}_{Q}=\frac{\hbar ^2}{2m_e} \boldsymbol{\nabla}\left( \frac{\nabla^{2}\sqrt{n}}{\sqrt{n}}\right). \label{quantum1}
\end{equation}
Equations (\ref{cont1}-\ref{quantum1}) have also been used to model superdense astrophysical bodies \cite{Jung} (i.e. the interior of Jupiter and massive white dwarfs, magnetars, and neutron
stars, etc.), intense laser-solid density plasma experiments \cite{Kremp}, and ultrasmall electronic devices \cite{Markowich} \cite{Shpatakovskaya}, carbon nanotubes \cite{Wei} and quantum diodes \cite{Ang}. Quantum hydrodynamical models have also revealed important features occurring in superfluidity \cite{Loffredo} and superconductivity \cite{Feynman}.\par
We now consider a cylindrical nanowire of radius $a$ and length $L\gg a$, in such a way that the system can be regarded as quasi one-dimensional along the longitudinal direction. In that case, we decompose the laplacian operator, which can be  written as $\nabla ^{2}=\nabla _{\perp }^{2}+\partial ^{2}/\partial _{z}^{2}$. Any relevant quantity $\Psi(=n, u, \phi)$ present in the above equations can therefore be decomposed as

\begin{equation}
\Psi (r,\theta ,z,t)=\sum\limits_{l,m}\Psi _{l,m}(z,t)J_{m}(k_{\perp;l,m}r)\exp (im\theta ),  \label{decomp}
\end{equation}
where $l$ and $m$ are integers ($\vert l \vert \leq m$), $k_{\perp;l,m}=\alpha _{l,m}/a$ is the transverse wavenumber and $\alpha _{l,m}$ stands for the $l$th zero of the Bessel function of order $m$. In the present work, it is considered the low-lying modes $l=0$ and $m=1$ only, without any loss of generality but for the matter of definiteness (for simplicity, we denote the transverse wavevector as $k_{\perp}$ in the remaining of the paper). After linearization, Eqs. (\ref{cont1}-\ref{poisson}) can be expressed in the following form 

\begin{equation}
\frac{\partial n_{l,m}}{\partial t}+n_{0}\frac{\partial u_{l,m}}{\partial z}=0,  \label{cont2}
\end{equation}

\begin{equation}
\left(\frac{\partial }{\partial t}+i\gamma \right) u_{l,m}=\frac{e}{m_e} \frac{\partial \phi _{l,m}}{\partial z}-\frac{v_{F}^{2}}{n_{0}}\frac{\partial n_{l,m}}{\partial z}+\frac{\hbar ^{2}}{4m_e^2n_0}\frac{\partial^3n_{l,m}}{\partial z^3},  \label{momentum2}
\end{equation}

\begin{equation}
\left(\frac{\partial ^{2}}{\partial _{z}^{2}}-k_{\perp}^{2}\right) \phi_{l,m}=\frac{e}{\varepsilon _{0}}\left(n_{l,m}\right).  \label{poisson2}
\end{equation}
After Fourier transforming eqs. (\ref{cont2}-\ref{momentum2}), and using the constitutive relation $\mathbf{D(\mathbf{k},\omega)}=\epsilon_{0}\epsilon(\mathbf{k},\omega)\mathbf{E(\mathbf{k},\omega)}$, we can easily derive the dielectric function for the conduction band

\begin{equation}
\epsilon(\mathbf{k},\omega)=1-\frac{k^2}{k^2+k_{\perp}^2}\frac{\omega_{p}^2}{\omega(\omega+i\gamma)-v_{F}^2k^2+h^2k^4/4m^2}.\label{diel2}
\end{equation} 
This expression describes a nonlocal, dispersive dielectric response of the system where both the effects of quantum diffraction ($\sim k^4$) and the finiteness of the system ($\sim k_{\perp}^2$) are taken into account and can also be used to described low-frequency electron waves in quantum plasma \cite {Hugo}. In the limit where such effects are negligible, one easily recovers the Drude model in Eq. (\ref{drude}).

\begin{figure}
\subfigure[]{
\includegraphics[scale=0.75]{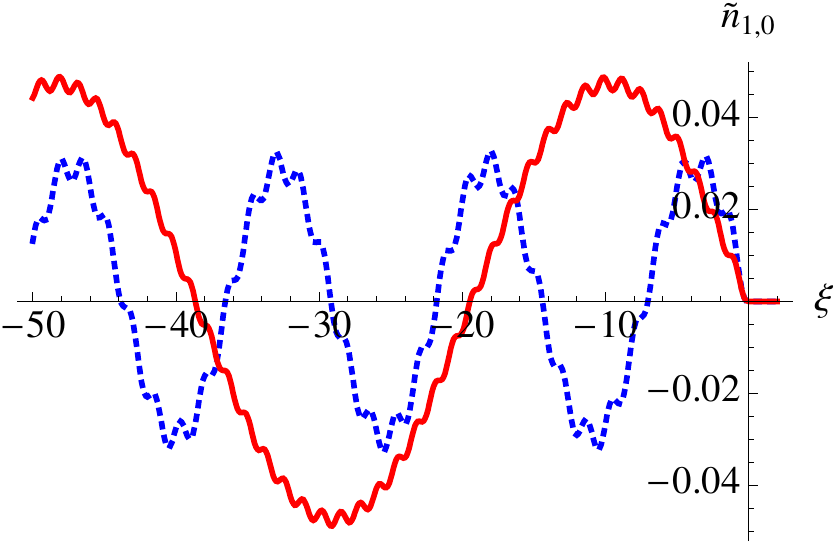}}
\subfigure[]{
\includegraphics[scale=0.75]{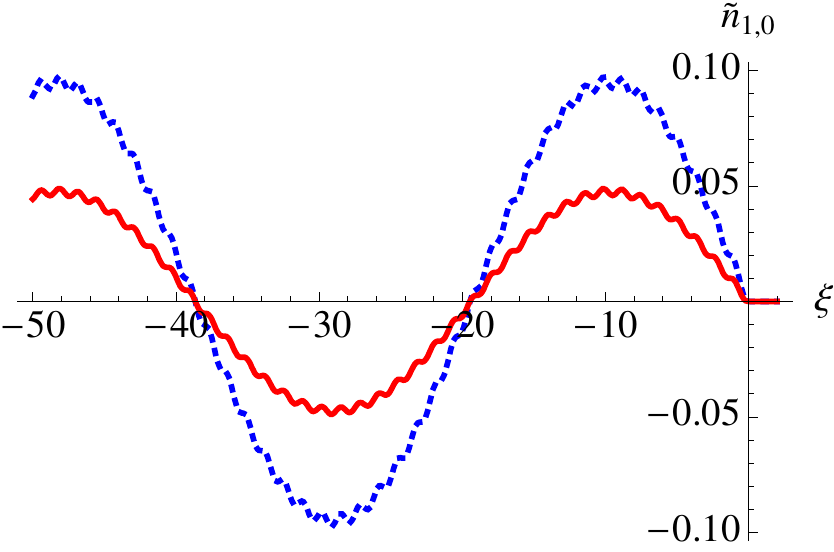}}
\caption{(Color online) The excitation of a wakefield due to a short electron pulse as function of: (a) the velocity of the pulse, $v_{0}=1.85 v_{F}$ (blue dashed
line), and $v_{0}=2.21 v_{F}$ (red solid line), for a width of $\sigma =0.1\lambda_{F}$; (b) the width of the electron pulse, $\sigma =0.1\lambda_{F}$ (blue dashed line) and $\sigma=0.2\lambda_{F}$ (red solid line), obtained for $v_{0}=2.21 v_{F}$.}
\label{fig:wakefield}
\end{figure}

We now examine the consequences of a propagation of an electron pulse in a dispersive medium with dielectric response given by Eq. (\ref{diel2}). In that case, the electron density is given by $n=n_{0}+n_{l,m}+N_{l,m}$, where $N_{l,m}$ is the electron pulse. Plugging into Eq. (\ref{poisson2}), 

\begin{equation}
\left( \frac{\partial ^{2}}{\partial {z}^{2}}-k_{\perp}^{2}\right) \phi_{l,m}=\frac{e}{\epsilon _{0}}\left(n_{l,m}+N_{l,m}\right), \label{poisson3}
\end{equation}
and assuming a typical situation where the collision frequency is negligible, $\gamma\ll\omega_{p}$ \cite{johnson}, Eqs. (\ref{cont2},\ref{momentum2},\ref{poisson3}) easily yield

\begin{eqnarray}
&&\left(\frac{\partial ^2}{\partial t^2}+\omega _p^2-v_F^2 \frac{\partial^2}{\partial z^2}+\frac{\hbar^2}{4m_e^2}\frac{\partial^4}{\partial z^4}\right) \notag \\
&&\times \left(\frac{\partial^2}{\partial z^2}-k_{\perp}^{2} \right) n_{l,m}+k_{\perp}^{2}\omega _{p}^{2}n_{l,m}  \label{pulse1}\\
&&=-\omega _{p}^{2}\frac{\partial^2}{\partial z^2}N_{l,m}.  \notag
\end{eqnarray}
Assuming that the electron pulse propagates with velocity $\mathbf{v}_{0}=v_{0}\hat{\mathbf{z}}$, we introduce the axial Lagrange coordinate $\zeta=z-v_{0}t$ to get

\begin{eqnarray}
&&\left\{ \frac{\partial ^{2}}{\partial \tau ^{2}}+(V^{2}-1)\frac{\partial^{2}}{\partial \xi ^{2}}-2V\frac{\partial ^{2}}{\partial \tau \partial \xi }+1\right. \notag\\
&& \left. +\frac{H^{2}}{4}\frac{\partial ^{4}}{\partial \xi ^{4}}\right\}\left( \frac{\partial ^{2}}{\partial \xi ^{2}}-K_{\perp l,m}^{2}\right) \tilde{n}_{l,m}+K_{\perp l,m}^{2}\tilde{n}_{l,m}  \label{pulse2} \\
&& =-\frac{\partial ^{2}}{\partial \xi ^{2}}\tilde{N}_{l,m},  \notag
\end{eqnarray}
where $\xi=\zeta/\lambda_{F}$, $\tau=\omega_{p}t$, $\tilde{N}_{l,m}=N_{l,m}/n_{0}$, $K_{\perp}=k_{\perp}\lambda _{F},$ and $V=v_{0}/v_{F}$ are dimensionless variables. Here, $H=\hbar \omega _{p}/(2k_{B}T_{F})$ is a dimensionless quantum. For short electron pulses, $\omega _{pe}^{-1}\gg\tau$, where $\tau$ is the typical duration of the pulse, an electrostatic wakefield is expected to be excited (for Au nanowires, the plasma frequency $\omega _{pe}^{-1}$ is of the order of few femtosecond).  The stationary solution in the moving frame, ($\partial /\partial \tau \rightarrow 0)$, corresponds to the case where the pulse propagates with negligible deformation and can be easily solved. In that case, we may write Eq. (\ref{pulse2}) as

\begin{equation}
\left(K_{a}^{2}\frac{\partial ^{2}}{\partial \xi ^{2}}+\frac{H^{2}}{4}\frac{\partial ^{4}}{\partial \xi ^{4}}+K_{b}^{4}\right) \tilde{n}_{l,m}(\xi )=-\tilde{N}_{l,m}(\xi ),  \label{pulse3}
\end{equation}
where $K_{a}^{2}=V^{2}-1-H^{2}K_{\perp}^{2}/4$ and $K_{b}^{4}=1+K_{\perp}^{2}(1-V^{2})$. Taking the Fourier transform of Eq. (\ref{pulse3}), one obtains 

\begin{equation}
\hat{n}_{l,m}(K)=-\hat{N}_{l,m}(K)\hat{G}(K),  \label{pulse4}
\end{equation}
where $\hat{G}(K)$ is the Fourier transformed Green function and can be expressed as 

\begin{equation}
\hat{G}(K)=\frac{1}{\left( K^{2}-K_{+}^{2}\right) \left(K^{2}-K_{-}^{2}\right) },  \label{green1}
\end{equation}
with
\begin{equation}
K_{\pm }^{2}=\frac{K_{a}^{2}\pm \sqrt{K_{a}^{4}-H^{2}K_{b}^{4}}}{H^{2}/2}. \label{green2}
\end{equation}
The solution to Eq. (\ref{pulse3}) is therefore readily obtained via the convolution theorem

\begin{equation}
\tilde{n}_{l,m}\left( \xi \right) =\int_{-\infty }^{\infty }\tilde{N}_{l,m}(\xi_{0})G(\xi -\xi _{0})d\xi _{0},  \label{solution1}
\end{equation}
where $G(\xi -\xi _{0})$ is the inverse Fourier transform of $G(K)$, which leads to the following solution to the density perturbation created by a short electron pulse moving along the axis of the nanowire

\begin{eqnarray}
&& \tilde n_{l,m}(\xi) =\frac{1}{K_+^2-K_-^2}\int_{-\infty}^{\infty }d\xi _0 \left( \frac{\sin K_+(\xi -\xi _0)}{K_+}\right. \notag \\
&& \left. -\frac{\sin K_-(\xi -\xi _0)}{K_-}\right) \Theta (\xi_{0}-\xi )\tilde{N}_{l,m}(\xi _{0}).  \label{solution2}
\end{eqnarray}
with $\Theta (x)$ representing the step function.  We can use $\tilde{N}_{l,m}(\xi_{0})=N_{0}\exp \left( -\xi _{0}^{2}/\sigma ^{2}\right)$ to describe a gaussian electron pulse of width $\sigma$. The later equation describes the excitation of a wakefield when a short pulse is set to propagate in a metallic nanowire. 

Stable electrostatic wakefields can be excited provided the inequalities $K_{a}^{2}>0,$ $K_{b}^{4}>0,$ and $K_{a}^{4}>H^{2}K_{b}^{4}, $ are simultaneously verified. These conditions constraint the velocity of the electron pulse as follows

\begin{equation}
\left(1+H-\frac{H^{2}K_{\perp}^{2}}{4}\right) ^{1/2}<V<\left(\frac{1+K_{\perp l,m}^{2}}{K_{\perp}^{2}}\right) ^{1/2}.  \label{stability1}
\end{equation}
We consider the concrete case of Au nanowires \cite{Manfredi}, for which the electron density is $n_{0}=5.85\times 10^{22}~\mathrm{cm}^{\mathrm{-3}}$ and the Fermi temperature is $T_{F}=63736.8$ K. In that case, we obtain $\omega _{p}\sim 1.54$ eV, $V_{F}\sim 1.39\times 10^{6}$ m/s and  $\lambda_{F}\sim 10.2$ nm. 

\begin{figure}
\includegraphics[scale=0.75]{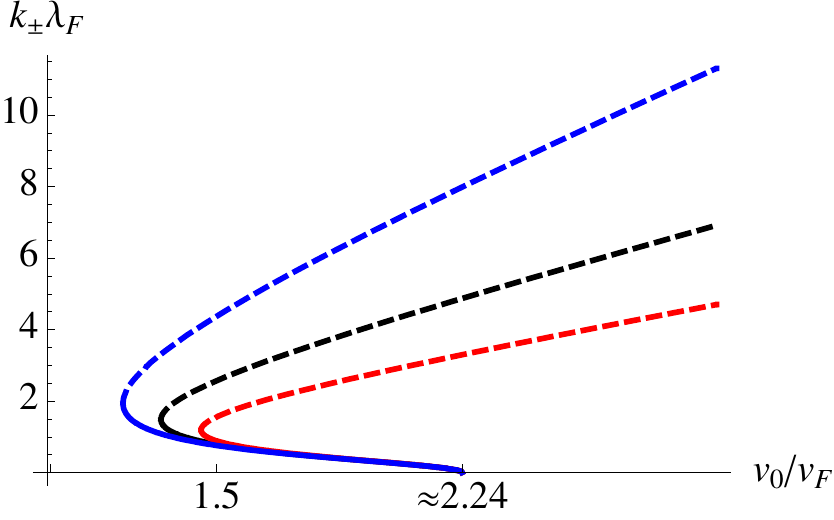}
\caption{(Color online) Wavenumbers $k_{-}$ (solid lines) and $k_{+}$ (dashed lines) for different values of the quantum parameter $H$, plotted for $k_{\perp}=0.5\lambda_{F}$: $H=0.51$ (blue line), $H=0.817$ (black line) and $H=1.2$ (red line). The lower cut-off corresponds to the upper limit in Eq. (\ref{stability1}) to the pulse velocity $v_{0}$ , which is independent of $H$. For $k_{\perp}\approx 2.41/a$, we have $v_{0}\approx 2.24 v_{F}$.}
\label{fig:parameter}
\end{figure}

The stability diagram of the wakefield is plotted in Fig. (\ref{fig:stability}) for $H\sim0.817$.  The critical value of the wavenumber, above which dynamical instability occurs, strongly depends on the quantum parameter $H$ (the ratio of the plasmon to the Fermi energies) and is defined as $k_{\perp}^{cr}=\sqrt{2/H}\sim 1.52/\lambda_{F}$. Stable wakefield solutions for different sets of parameters are illustrated in Fig. (\ref{fig:wakefield}). It is observed that the quantum oscillations, of periodicity $k_{+}$, are enhanced for very short pulses, i.e. $k_{+}\sigma \lesssim 1$. The pulse velocity $v_{0}$ also plays a role on the amplitude of the wakefield, as it is related with the value of $k_{-}$ and $k_{+}$. The width of pulse also seems to change the amplitude of the wakefield and that can be easily understood by noticing that Eq. (\ref{solution2}) provides a factors of the form $\exp(-k_{\pm}\sigma)$, which damps the amplitude of the perturbation for long pulses. The difference between the modes $k_{\pm}$ can be enhanced by changing the value of the quantum parameter $H$ (see Fig. (\ref{fig:parameter})). In practice, this simply corresponds to choose a different metal, since $H$ only depends on the nature of the sample. 
  
\begin{figure}
\includegraphics[scale=0.75]{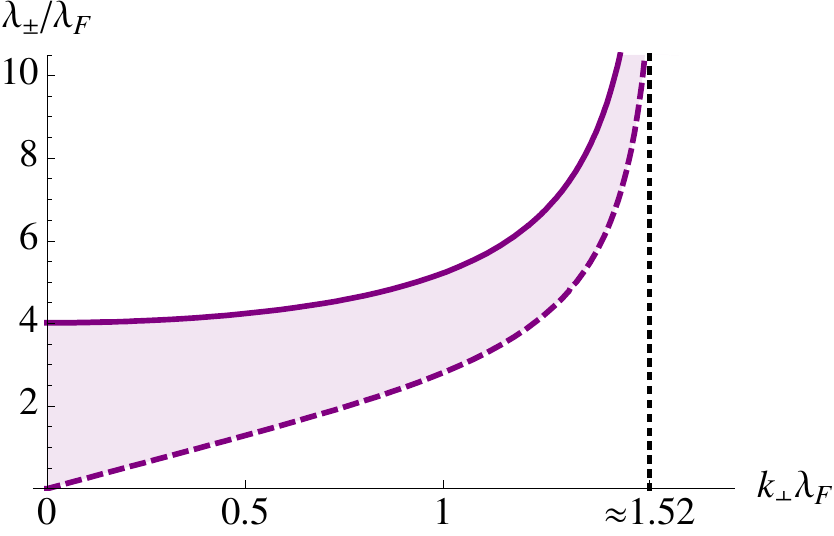}
\caption{(Color online) Accessible wavelength range (shadowed area) for the radiation emitted by the wakefield as function of transverse radius of the nanowire ($k_{\perp}\approx 2.41/a$, see discussion in the text). The lower and upper curves correspond to the boundaries of stable wakefields in Eq. (\ref{stability1}). The values fit the XUV wavelength range.}
\label{fig:radiation}
\end{figure}

The excitation of wakefields may also open the door for the investigation of possible new sources of radiation in nanoscale devices. The electron current created by the wakefield emit radiation at frequencies $\omega_{\pm}\sim ck_{\pm}$. In Fig. (\ref{fig:radiation}), we evaluate the range of frequencies generated by a short pulse propagating in a nanowire by plugging the stability condition (\ref{stability1}) into the definition of $k_{+}$ and $k_{-}$. For the special case of Au nanowires aforementioned, typical experiments are performed with cylinders of radius $a \sim 5$ nm, which corresponds to a transverse cut-off $k_{\perp}\sim 0.5/\lambda_{F}$. Stable structures can therefore modulate electrons at wavelengths of $\lambda \sim 1-4$ nm. The emission range can be broader or narrower provided that the value of $H$ is decreased of increased, respectively (see Fig. (\ref{fig:parameter})). When properly amplified, these signals can be used to produce radiation in the extreme ultra-violet (XUV) range. This may be particularly interesting in experiments where high-sensitivity, low-power XUV radiation is needed. 

To summarize, we have extended the usual hydrodynamical description of metallic nanowires and derived an expression for the nonlocal (dispersive) dielectric constant accounting for both the quantum diffraction of the electrons and the finite size effects in the transverse direction. This dispersive dielectric response of the system is then considered to investigate the excitation of wakefields due to the propagation of a short electron pulse. We show that stable wakefield generation is expected to occur for a reasonable set of parameters experimentally accessible. It was also shown that the competition between the classical and quantum behaviors of the system provide two scales of electronic modulation ($k_{+}$ and $k_{-}$), which are related via the finiteness condition along the radial direction. The existence of this competition (which is a consequence oh the dispersive dielectric properties of nanowires) has also been pointed out by other authors \cite{Shukla}. Finally, we investigate a possible application of wakefield excitation in nanowires as a source of radiation in the XUV range. With the available state-of-the-art technology, very short pulses can be generated also at low power (using, e.g., lasers diodes \cite{Rafailov}), allowing wakefields to be studied outside the traditional areas of plasma physics and, in particular, the investigation of wakefield phenomena in the context of nanotechnology.

The authors SA and HT acknowledge the financial support from Funda\c c\~ao o para a Ci\^{e}ncia e Tecnologia (FCT), Lisbon-Portugal through the grants SFRH/BPD/63669/2009 and SFRH/BD/37452/2007.

\end{document}